\documentclass[reprint,superscriptaddress,amsmath,amssymb,aps,nofootinbib]{revtex4-2}
\usepackage{graphicx}
\usepackage{dcolumn}
\usepackage{bm}
\usepackage[colorlinks,linkcolor=blue,citecolor=blue]{hyperref}

\begin{document}

\title{Universality of Osmotic Equation of State in Star Polymer Solutions}

\author{Takashi~Yasuda}
\affiliation{Graduate School of Engineering, The University of Tokyo, 7-3-1 Hongo, Bunkyo-ku, Tokyo, Japan.}

\author{Masanobu~Ino}
\affiliation{Graduate School of Engineering, The University of Tokyo, 7-3-1 Hongo, Bunkyo-ku, Tokyo, Japan.}

\author{Takamasa~Sakai}
\email[Corresponding author: ]{sakai@gel.t.u-tokyo.ac.jp}
\affiliation{Graduate School of Engineering, The University of Tokyo, 7-3-1 Hongo, Bunkyo-ku, Tokyo, Japan.}

\author{Naoyuki~Sakumichi}
\email[Corresponding author: ]
{sakumichi@gel.t.u-tokyo.ac.jp}
\affiliation{Graduate School of Engineering, The University of Tokyo, 7-3-1 Hongo, Bunkyo-ku, Tokyo, Japan.}

\date{\today}

\begin{abstract}
We experimentally measure the osmotic pressures of linear polymers and three-, four-, and eight-arm star polymers in a good solvent via membrane osmometry.
These results reveal that the osmotic equations of state in the star polymer solutions are universally described by the same scaling function that describes linear polymer solutions.
This universality is achieved by canceling increasing overlap concentrations and decreasing osmotic pressure, owing to the increased arm number.
We further clarify the molar mass and arm number dependencies of the gyration radius and interpenetration factor, ensuring universality in star polymer solutions.
\end{abstract}

\maketitle

Star polymers consisting of several linear chains radiating from a central core have attracted considerable attention in recent decades for numerous applications in life sciences and nanotechnology~\cite{daoud1982star,wu2015star,ren2016star}.
Recently, the value of the use of ``semidilute'' star polymer solutions, where polymers overlap, has increased. 
This is because gels with a precisely controlled polymer network~\cite{nakagawa2022star} can be synthesized by end-linking star polymers in a semidilute regime; that is, the polymer concentration $c$ exceeds the overlap concentration $c^*$. 
Chemical gels~\cite{sakai2008design, li2019polymer,huang2020simple,fujiyabu2022tri} and physical gels~\cite{grindy2015control,yesilyurt2017mixed,marco2020linking,Ahmadi2021,chen2021control,ohira2022star} consisting of precisely controlled permanent and transient networks, respectively, have been fabricated, with their reproducibility and controllability leading to significant advances in polymer physics~\cite{Zhong2016,Seiffert2017,yoshikawa2021negative,fujiyabu2021temperature,sakumichi2021linear,sakumichi2022semidilute}.

Although providing insights into semidilute star polymer solutions ($c>c^*$) is becoming increasingly important, the physical understanding is limited.
For example, the governing law of their osmotic pressure $\Pi$, one of the most fundamental physical properties, remains unclear despite pioneering efforts~\cite{higo1983osmotic,cherayil1986osmotic,adam1991concentration,merkle1993osmotic,roovers1995thermodynamic,burchard1999solution}.
The major difficulty is that reliable measurement (through membrane osmometry) of $\Pi$ of semidilute star polymer solutions is reported only for three-arm star polymers~\cite{higo1983osmotic}.
This situation is in contrast to dilute star polymer solutions ($c<c^*$); the gyration radius and second virial coefficient have been extensively investigated via light scattering~\cite{grest1987structure,ohno1996monte,rubio1996monte,di2002monte,hsu2004scaling,ida2008monte,sato1987second,nakamura1991third,miyaki1978excluded,okumoto1998excluded,okumoto2000excluded,yamamoto1971more,fukuda1974solution,roovers1980hydrodynamic,roovers1986linear,douglas1990characterization,roovers1974preparation,bauer1989chain,khasat1988dilute,huber1984dynamic,roovers1983analysis}.

\begin{figure}[t!]
\centering
\includegraphics[width=\linewidth]{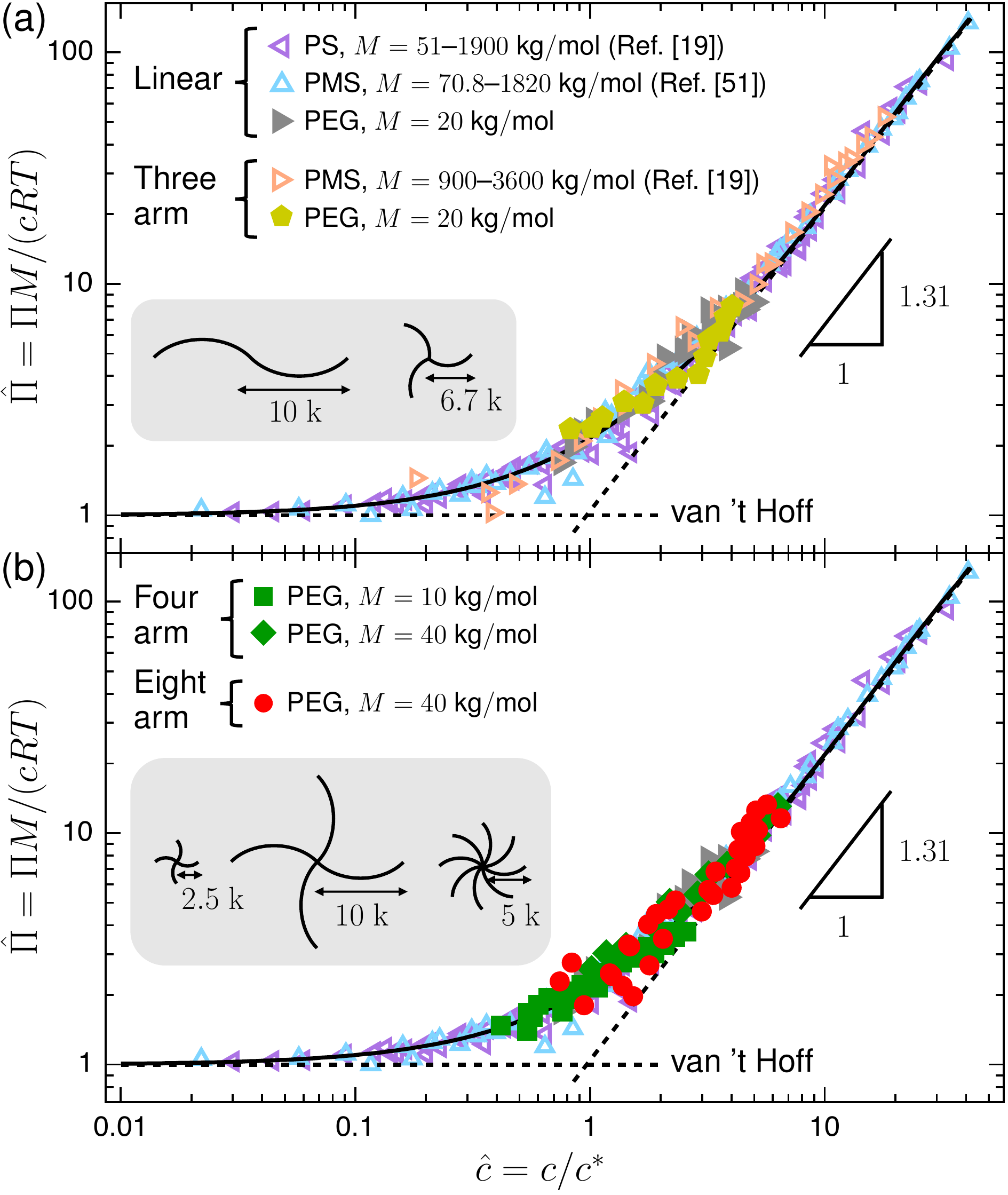}
\caption{
Universality of osmotic EOS in linear and star polymers in good solvents.
The open triangles represent two kinds of linear polymers 
[poly($\alpha$-methylstyrene) (PMS) with $M=70.8$--$1820$ kg$/$mol~\cite{noda1981thermodynamic} and poly(styrene) (PS) with $M=51$--$1900$ kg$/$mol~\cite{higo1983osmotic}] in toluene solutions.
The open stars represent the three-arm star polymers [PMS with $M=900$--$3600$ kg$/$mol~\cite{higo1983osmotic}] in toluene solutions.
The gray-filled triangles represent the linear polymers [poly(ethylene glycol) (PEG) with $M=20$ kg$/$mol] in aqueous solutions.
The yellow pentagons, green squares and diamonds, and red circles represent the star PEG in aqueous solutions for (a) three arms with $M=20$ kg$/$mol, (b) four arms with $M=10$ and $40$ kg$/$mol and eight arms with $M=40$ kg$/$mol, respectively.
We reported the data for four arms in Ref.~\cite{yasuda2020universal}.
All data collapse onto the universal EOS given by Eq.~(\ref{eq:EOS}) (black solid curve), which is asymptotic to the van 't Hoff law ($\hat{\Pi}=1$) as $\hat{c}\to 0$ and the scaling law in Eq.~(\ref{eq:semidilute}) as $\hat{c}\to \infty$ (black dashed lines).
}
\label{fig:EOS}
\end{figure}

Before considering star polymers, we briefly review $\Pi$ of (electrically neutral) flexible linear polymer solutions.
Numerous theoretical~\cite{des1975lagrangian,de1979scaling,ohta1982conformation,ohta1983theory,oono1985statistical} and experimental~\cite{noda1981thermodynamic,des1982osmotic,higo1983osmotic,noda1984semidilute} studies have clarified that $\Pi$ of dilute and semidilute linear polymers in good solvents is \textit{universally} described by the following osmotic equation of state (EOS):
\begin{equation}
\hat{\Pi} = f_\mathrm{\Pi}\left({\hat{c}}\right),
\label{eq:EOS}
\end{equation}
where $\hat{\Pi}\equiv\Pi M/(cRT)$ is the reduced osmotic pressure and $\hat{c}\equiv c/c^{*}$ is the reduced polymer mass concentration $c$ normalized by overlap concentration $c^{*}\equiv1/(A_{2}M)$.
Here, $M$, $R$, $T$, and $A_{2}$ are the molar mass, gas constant, absolute temperature, and second virial coefficient, respectively.
Figure~\ref{fig:EOS}(a) demonstrates that different types of linear polymer solutions (open triangles) converge to Eq.~(\ref{eq:EOS}) (black solid curves)~\cite{noda1981thermodynamic,higo1983osmotic}.
In the dilute regime ($0\leq \hat{c} <1$), each polymer chain is isolated, and $\Pi$ is described through the virial expansion~\cite{flory1953principles}:
\begin{equation}
\hat{\Pi} = 1 + \hat{c} + \gamma\,\hat{c}^{2}+ \cdots,
\label{eq:sqr}
\end{equation}
where $\gamma \approx 0.25$ is the dimensionless virial ratio of the third virial coefficient~\cite{stockmayer1952third}.
In the semidilute regime ($\hat{c}>1$), the polymer chains overlap without isolation, and $\Pi$ is independent of $M$.
Consequently, Eq.~(\ref{eq:EOS}) is asymptotic to the following scaling law [black dashed line in Fig.~\ref{fig:EOS}(a)]~\cite{des1975lagrangian,de1979scaling}:
\begin{equation}
\hat{\Pi} = K\hat{c}^{\frac{1}{3\nu -1}},
\label{eq:semidilute}
\end{equation}
where $K\approx 1.1$ and $1/(3\nu-1)\approx1.31$.
Here, $\nu\approx0.588$ is the excluded volume parameter for good solvents~\cite{flory1953principles} and is known as the universal critical exponent of the self-avoiding-walk universality class~\cite{pelissetto2002critical}.

In this Letter, we investigate the universal law of $\Pi$ in star polymer solutions with up to eight arms.
We experimentally measure $\Pi$ via membrane osmometry and
find that $\Pi$ depends on the number of arms $f$ but not on the molar mass $M$ in the semidilute regime ($c>c^*$).
Furthermore, we find that the dimensionless $\hat{\Pi}$ of star polymer solutions is described by the same EOS as the linear polymer solutions [Eq.~(\ref{eq:EOS})] in the dilute and semidilute regimes. 
Figure~\ref{fig:EOS} demonstrates that the linear, three-arm, four-arm, and eight-arm star polymer solutions agree with Eq.~(\ref{eq:EOS}) (black solid curve).
The difference between the linear ($f=2$) and star ($f\geq 3$) polymer solutions manifests as increasing $c^*$ and decreasing $\Pi$, owing to increased $f$.
Furthermore, we clarify $M$ and $f$ dependences of the gyration radius $R_{g}$ ($\propto M^{\nu}$) and interpenetration factor $\Psi^{*}$, ensuring the universality of EOS in star polymer solutions.\\

\textit{Materials and methods}.\,---
We used linear, three-arm, four-arm, and eight-arm star PEG with $M=20$ kg$/$mol (Polyethylene glycol 20000, Sigma-Aldrich), $M=20$ kg$/$mol (SUNBRIGHT GL2-200MA, NOF Co., Japan), $M=10$ and $40$ kg$/$mol (SUNBRIGHT PTE-100MA and 400MA, NOF Co., Japan), and $M=40$ kg$/$mol (8-arm PEG-OH(TP), XIAMEN SINOPEG BIOTECH Co., Ltd., China), respectively.
We confirmed that the effect of the difference in end-functional groups on osmotic pressure is negligible [Supplemental Material (SM), Sec.~S1].
We prepared samples for the initial polymer mass concentrations $c_0=20$--$160$ g$/$L in aqueous solutions at $298$~K.
Here, we define the polymer mass concentration as the polymer weight divided by the solvent volume, rather than the solution volume, to extend the universality of EOS to higher concentrations (see SM, Sec.~S1 in Ref.~\cite{yasuda2020universal}).

\begin{figure*}[t!]
\centering
\includegraphics[width=\linewidth]{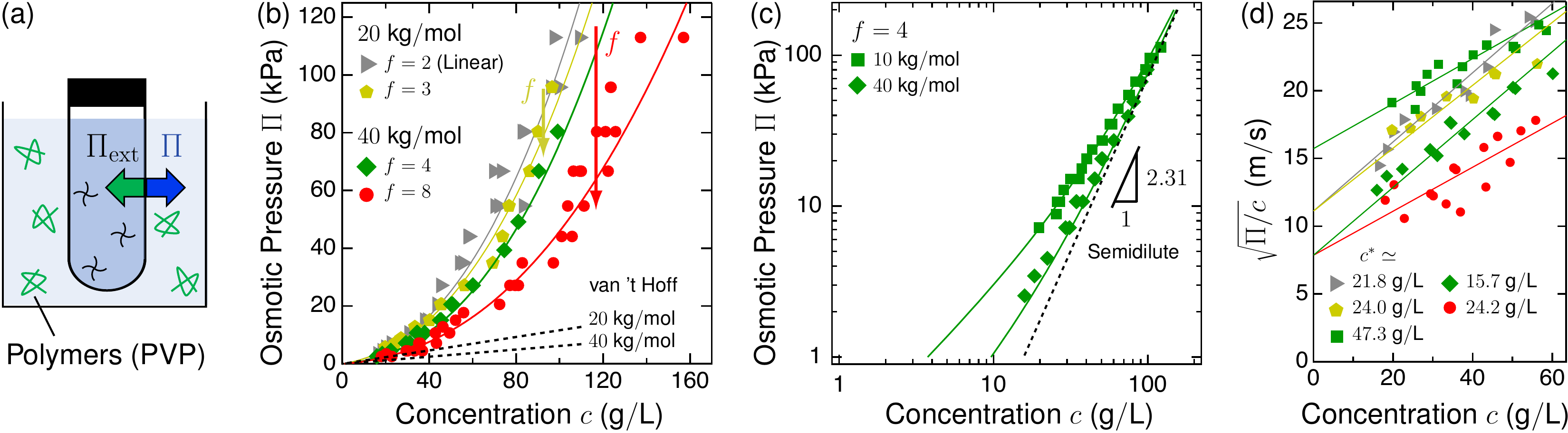}
\caption{
Osmotic pressure $\Pi$ of aqueous solutions with the linear and star PEG.
(a) Membrane osmometry with a microdialyzer in external aqueous polymer (PVP) solutions. 
We can determine $\Pi$, using $\Pi=\Pi_\mathrm{ext}$ at equilibrium.
(b) Dependence of $\Pi$ on polymer mass concentration $c$ for arm numbers $f=2$ (linear, gray triangles) and $f=3$ (yellow pentagons) with the molar mass $M=20$ kg$/$mol, and $f=4$ (green diamonds) and $f=8$ (red circles) with $M=40$ kg$/$mol.
In the dilute limit ($c\to 0$), $\Pi$ follows the van 't Hoff law, $\Pi=cRT/M$ (black dashed lines).
(c) Dependence of $\Pi$ on $c$ with $M=10$ and $40$ kg$/$mol for $f=4$.
In the semidilute regime ($c>c^{*}$), $\Pi$ is asymptotic to the semidilute scaling law given by Eq.~(\ref{eq:semidilute}) with the exponent $3\nu/(3\nu-1)\approx2.31$ (black dashed line).
(d) Square-root plot of the osmotic pressure.
Each solid line is a least-square fit of Eq.~(\ref{eq:squre-root-plot}), where the fit parameter is $c^*$. 
In (b) and (c), each solid curve represents the universal EOS given by Eq.~(\ref{eq:EOS}), using $c^{*}$ evaluated in (d) with each $M$.
}
\label{fig:osmotic}
\end{figure*}

We measured $\Pi$ of the linear and star polymer solutions using membrane osmometry.
We used controlled aqueous poly(vinylpyrrolidone) (PVP, K90, Sigma-Aldrich) solutions at $T=298$~K, whose concentration dependence of osmotic pressure $\Pi_\mathrm{ext}$ was reported as
$\Pi_\mathrm{ext} = 21.27c_\mathrm{ext} + 1.63c_\mathrm{ext}^{2} + 0.0166c_\mathrm{ext}^{3}$
for $c_\mathrm{ext} \leq 200$ g$/$L~\cite{vink1971precision}. 
Here, the PVP concentration $c_\mathrm{ext}$ is defined as the polymer weight divided by the solution volume.
As shown in the schematic in Fig.~\ref{fig:osmotic}(a), each solution sample was placed in a microdialyzer (MD300, Scienova) with a semipermeable membrane (mesh size, $3.5$~kDa).
Each microdialyzer was immersed in an aqueous polymer (PVP) solution at a certain concentration $c_\mathrm{ext}$ with stirring.
Subsequently, each system reached equilibrium at $\Pi=\Pi_\mathrm{ext}$.
(Reaching equilibrium was ensured. See SM, Sec.~S3.)
Each solution sample was changed in weight ($W_{0} \to W$) and polymer mass concentration ($c_{0} \to c$) from the as-prepared state to equilibrium.
For each sample, we tuned $c_\mathrm{ext}$ to satisfy $W/W_{0}\approx1$.
We calculated $c$ as 
$c=c_{0}/\left[W/W_{0}+(W/W_{0}-1)c_{0}/\rho_{s}\right]$,
where $\rho_{s}$ is the density of solvents, using $\rho_\mathrm{water}\approx1.0\times10^3$ kg$/$m$^{3}$ for aqueous solvents.
All experimental results for $\Pi$ are available in SM, Sec.~S3.\\

\textit{Results}.\,---
Figure~\ref{fig:osmotic}(b) presents a comparison of the $c$ dependence of $\Pi$ for different arm numbers ($f$), demonstrating that $\Pi$ decreases with increasing $f$ when $M$ is constant.
To compare the samples of different $f$ at fixed $M$, we disregarded the polydispersity ($\leq 1.2$) of each sample and slight differences in the average molar mass with different $f$ due to synthesis inaccuracies.
For a low $c$ (dilute regime), $\Pi$ follows $\Pi=cRT/M$ (van 't Hoff law), which is independent of $f$.
By contrast, for a high $c$ (semidilute regime), $\Pi$ decreases with increasing $f$ (red and yellow arrows) at fixed $c$ and $M$.
The decrease in $\Pi$ among $f=4$ and $8$ (red arrow) is significant, whereas only minor differences arise between $f=2$ and $3$ (yellow arrow).
Consequently, the decrease in $\Pi$ was not recognized in Ref.~\cite{higo1983osmotic} examining $f=2$ and $3$.
This decrease in $\Pi$ owing to branching cannot be explained by the conventional scaling argument [described before Eq.~(\ref{eq:semidilute})] for semidilute linear polymer solutions; the origin of this decrease is discussed below.

Figure~\ref{fig:osmotic}(c) shows the $c$ dependence of $\Pi$ with different $M$ for four-arm star polymer solutions, which are asymptotic to an $M$-independent scaling law in the semidilute regime ($c \gg c^{*}$).
The star polymer solutions exhibit the $M$-independent scaling law in the semidilute regime, showing the same characteristics as linear polymer solutions [Eq.~(\ref{eq:semidilute})]~\cite{des1975lagrangian,de1979scaling}.
The exponent of the scaling law $3\nu/(3\nu-1)\approx2.31$ is independent of $f$, whereas the prefactor of $\Pi$ depends on $f$.

We show that $c^{*}$ at fixed $M$ increases as $f$ increases by evaluating $c^{*}$ from the square-root plots~\cite{flory1953principles} of the experimental results shown in Fig.~\ref{fig:osmotic}(b) and (c).
From the virial expansion in Eq.~(\ref{eq:sqr}), we have 
$\hat{\Pi}
 = \left[1+\hat{c}/2
 + \left(\gamma-1/4 \right)\hat{c}^2/2
 \right]^2
 + O\left(\hat{c}^3\right)
$.
Together with $\gamma\approx1/4$ for star polymer solutions with few arms (SM, Sec.~S4), we have
\begin{equation}
\sqrt{\frac{\Pi}{c}}
\simeq \sqrt{\frac{RT}{M}}
\left(1+\frac{c}{2c^{*}}\right)
\label{eq:squre-root-plot}
\end{equation}
for a small $c/c^{*}$.
Thus, $c^{*}$ for each sample can be estimated by the slopes of the best-fit lines in Fig.~\ref{fig:osmotic}(d).
We obtained $c^{*}=21.8(7)$~g$/$L for $f=2$ with $M=20$~kg$/$mol; $c^{*}=24.0(10)$~g$/$L for $f=3$ with $M=20$~kg$/$mol;
$c^{*}=47.3(21)$ and $15.7(4)$~g$/$L for $f=4$ with $M=10$ and $40$~kg$/$mol, respectively;
and $c^{*}=24.2(13)$~g$/$L for $f=8$ with $M=40$~kg$/$mol.
Values in parentheses represent standard errors.
The obtained $c^{*}$ increases as $f$ increases at fixed $M$.

\begin{figure}[b!]
\centering
\includegraphics[width=\linewidth]{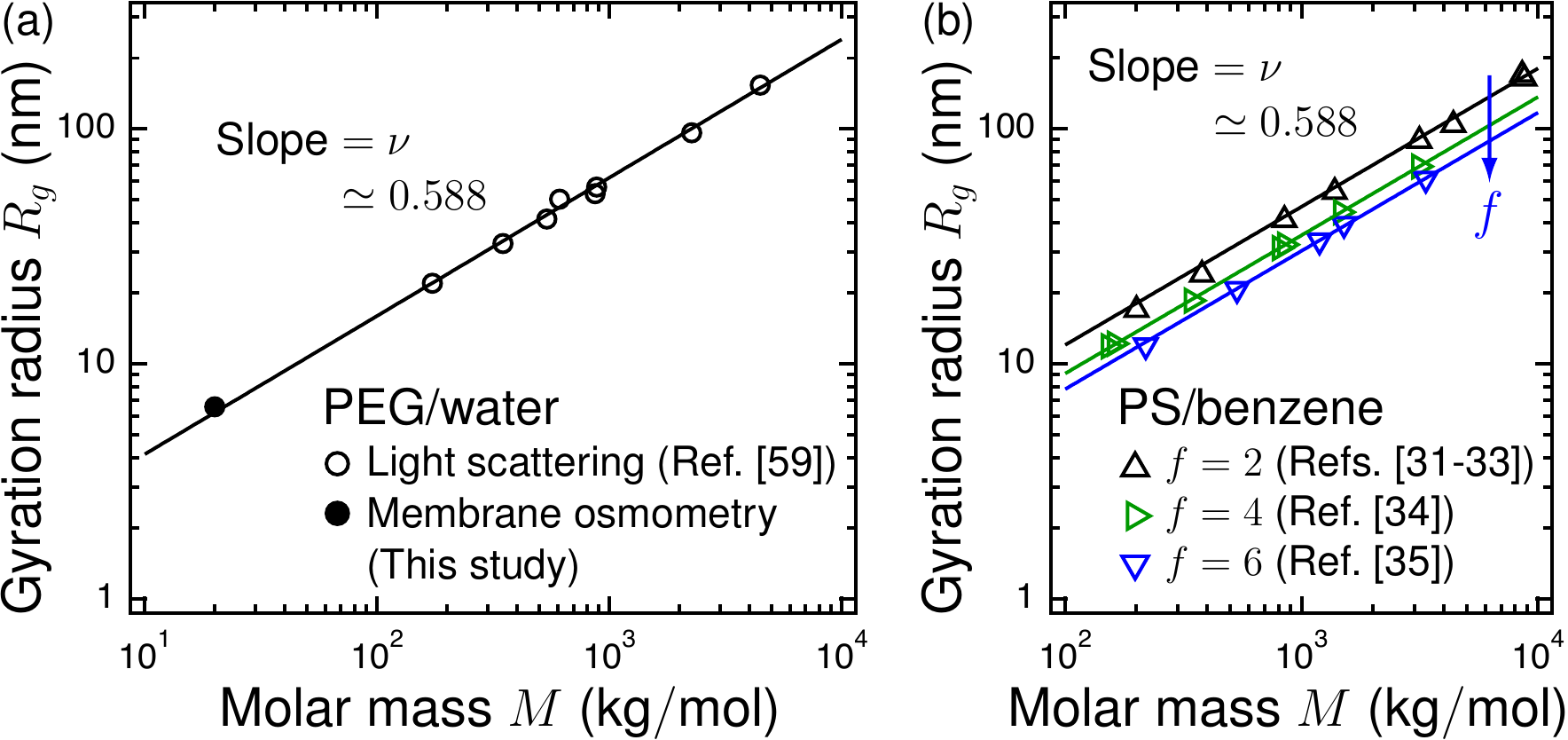}
\caption{
Molar mass ($M$) dependence of gyration radius $R_{g}$ at $298$~K for (a) linear PEG in aqueous solutions and (b) linear (black upward triangles), four-arm (green rightward triangles), and six-arm (blue downward triangles) poly(styrene) in benzene solutions.
Each solid line represents $R_{g}\sim M^\nu$ with $\nu\approx0.588$.
In (a), the results measured via membrane osmometry (filled circle) are consistent with those measured via light scattering~\cite{kawaguchi1997aqueous} (open circles).
In (b), the data are taken from Refs.~\cite{sato1987second,nakamura1991third,miyaki1978excluded,okumoto1998excluded,okumoto2000excluded}, measured via light scattering.
}
\label{fig:Rg}
\end{figure}

\begin{figure*}[t!]
\centering
\includegraphics[width=\linewidth]{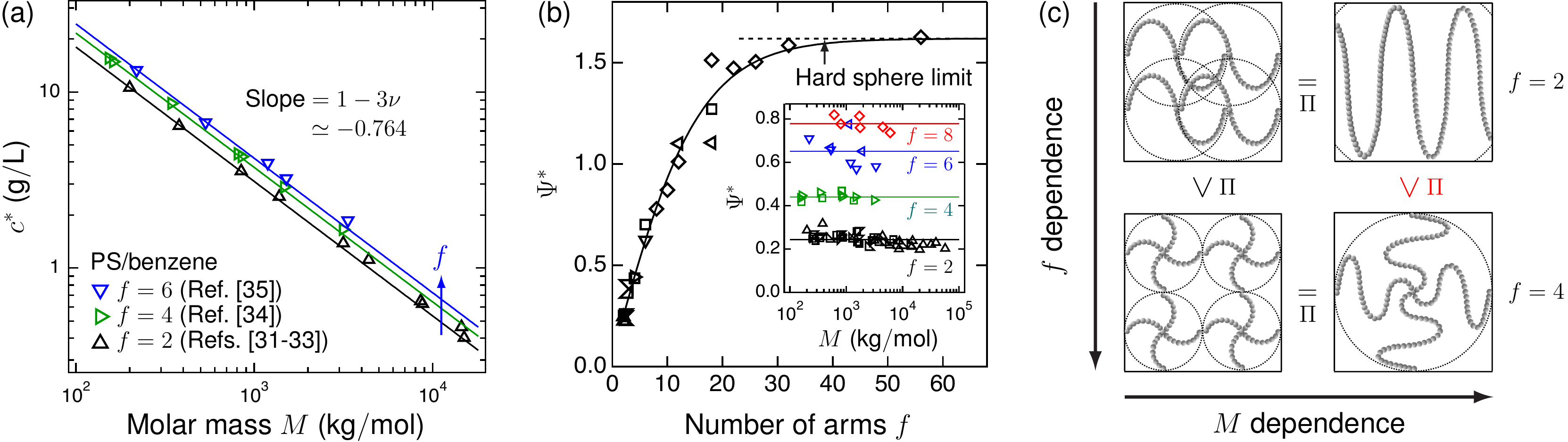}
\caption{
(a) Molar mass dependence of $c^*$ for the linear (black upward triangles), four-arm (green rightward triangles), and six-arm (blue downward triangles) poly(styrene) in benzene solutions at $298$~K.
(b) Interpenetration factor $\Psi^{*}$ for various $f$.
The error bars indicate the standard deviation.
For large $f$, $\Psi^{*}$ is asymptotic to the hard-sphere value $\Psi^{*}\approx1.61$ (dashed line)~\cite{yamakawa1971modern}.
Inset: Molar mass dependence of $\Psi^{*}$ for $f=2$ (black symbols), $f=4$ (green symbols), $f=6$ (blue symbols), and $f=8$ (red symbols) with various polymers and solvents at different temperatures.
Here, different symbols indicate differences in the references.
The data in (a) and (b) are taken from Refs.~\cite{sato1987second,nakamura1991third,miyaki1978excluded,okumoto1998excluded,okumoto2000excluded} and Refs.~\cite{sato1987second,nakamura1991third,miyaki1978excluded,okumoto1998excluded,okumoto2000excluded,yamamoto1971more,fukuda1974solution,roovers1980hydrodynamic,roovers1986linear,douglas1990characterization,roovers1974preparation,bauer1989chain,khasat1988dilute,huber1984dynamic,roovers1983analysis}, respectively, measured via light scattering.
(c) Schematic of the $M$ and $f$ dependencies of $\Pi$ in the semidilute linear and star polymer solutions at fixed $c$.
}
\label{fig:previous}
\end{figure*}

We validate that the obtained $c^{*}\equiv 1/(A_{2}M)\approx21.8$~g$/$L for the linear PEG in aqueous solution is consistent with the gyration radius $R_{g}$ measured via light scattering in Ref.~\cite{kawaguchi1997aqueous}.
We use the following relation:
\begin{equation}
c^{*}=\frac{M}{4\pi^{\frac{3}{2}} R_{g}^{3} N_{A}\Psi^{*}},
\label{eq:c-star}
\end{equation}
where $N_{A}$ is the Avogadro constant and $\Psi^{*}$ is the interpenetration factor.
Because $\Psi^{*}\approx0.24$ for linear polymers~\cite{sato1987second,nakamura1991third,miyaki1978excluded,yamamoto1971more,fukuda1974solution,roovers1980hydrodynamic,roovers1986linear,douglas1990characterization}, we obtain $R_{g}=[M/(4\pi^{3/2}N_{A}\Psi^{*}c^{*})]^{1/3}=6.58(7)$~nm for $c^{*}=21.8(7)$~g$/$L.
The samples of $M=10$--$40$~kg$/$mol used in this study correspond to polymerization degrees $N \approx 227$--$909$, which are sufficiently large 
as $\Psi^{*}$ remains constant~\cite{rubio1996monte,ida2008monte}.
Figure~\ref{fig:Rg}(a) shows that $R_{g}$ reported in Ref.~\cite{kawaguchi1997aqueous} and $R_{g}$ measured in this study agree with the scaling law $R_{g} \propto M^{\nu}$ (solid line), with $\nu \approx 0.588$~\cite{pelissetto2002critical}.
This result validates $c^{*}$ determined in our analysis.

Using $\Pi$ shown in Fig.~\ref{fig:osmotic}(b) and (c) and $c^{*}$ evaluated in Fig.~\ref{fig:osmotic}(d), we demonstrate that the osmotic EOS in star polymer solutions with up to eight arms is universally described by Eq.~(\ref{eq:EOS}).
We rescaled the state variables from dimensional ($c$ and $\Pi$) to dimensionless ($\hat{c}$ and $\hat{\Pi}$) for each sample and yielded the filled symbols shown in Fig.~\ref{fig:EOS}.
The open symbols for linear and three-arm polymers in Fig.~\ref{fig:EOS} were obtained by reanalyzing the reported data~\cite{noda1981thermodynamic,higo1983osmotic} in the same manner, using $c^* \equiv 1/(A_2 M)$.
All symbols, including linear, three-arm, four-arm, and eight-arm polymers with various $M$, demonstrate the universality of the osmotic EOS given by Eq.~(\ref{eq:EOS})~\cite{noda1981thermodynamic,higo1983osmotic}.
This universality originates from canceling the decrease in $\Pi$ [Fig.~\ref{fig:osmotic}(b)] and the increase in $c^{*}$ [Fig.~\ref{fig:osmotic}(d)] as $f$ increases.

We emphasize that the $A_2$-based definition of the overlap concentration, $c^* \equiv 1/(A_2M)$, is essential in exhibiting the universality of osmotic EOS in star polymer solutions.
The commonly used $R_g$-based definition of the overlap concentration is $c^*_g \equiv 3M/(4\pi N_A R_g^3)$, where the polymer chains begin to overlap to fill the space.
However, $c/c^*_g$ is not the universal scaling parameter for star polymers, because $c^*_g/c^*=3\sqrt{\pi}\Psi^*$ includes the interpenetration factor $\Psi^*$.
Here, $\Psi^*$ depends on $f$ due to the high segment density near the central core preventing polymer chain penetration.
In Ref.~\cite{higo1983osmotic}, the universality of the osmotic EOS was not revealed because $c/c^*_g$ was used to compare $f=2$ and $f=3$.\\

\textit{Discussion}.\,---
To ensure the universality of Eq.~(\ref{eq:EOS}) (\textit{i.e.}, the function $f_\Pi$) for various polymers and solvents, we investigated $M$ and $f$ dependencies of (i) $c^*$ and (ii) $\Pi$.
Figure~\ref{fig:osmotic}(d) and (b) experimentally show (i) and (ii), respectively, for PEG in aqueous solutions;
$c^{*}$ is an increasing sequence of $f$, and $\Pi$ is a decreasing sequence of $f$.

Regarding (i), using the data reported in Refs.~\cite{sato1987second,nakamura1991third,miyaki1978excluded,okumoto1998excluded,okumoto2000excluded}, Figs.~\ref{fig:Rg}(b) and \ref{fig:previous}(a) show that $f$-arm star poly(styrene) in benzene solutions exhibits the scaling law $R_{g}\propto M^{\nu}$ and $c^*\propto M^{1-3\nu}$, respectively.
Here, the latter scaling law comes from Eq.~(\ref{eq:c-star}) with the fact that $\Psi^{*}$ is independent of $M$ (see below): $c^*\propto MR_{g}^{-3}\propto M^{1-3\nu}$.
Moreover, Figs.~\ref{fig:Rg}(b) and \ref{fig:previous}(a) show that $R_g$ and $c^*$ are decreasing and increasing sequences of $f$ at fixed $M$, respectively (blue arrows);
this is because each arm length is proportional to $M/f$.

Organizing the data of $R_g$ and $A_2$ obtained from Refs.~\cite{sato1987second,nakamura1991third,miyaki1978excluded,okumoto1998excluded,okumoto2000excluded,yamamoto1971more,fukuda1974solution,roovers1980hydrodynamic,roovers1986linear,douglas1990characterization,roovers1974preparation,bauer1989chain,khasat1988dilute,huber1984dynamic,roovers1983analysis}, Fig.~\ref{fig:previous}(b) shows that $\Psi^{*}$ is an increasing sequence of $f$ from $\Psi^{*}\approx 0.24$ for $f=2$ (linear polymer~\cite{sato1987second,nakamura1991third,miyaki1978excluded,yamamoto1971more,fukuda1974solution,roovers1980hydrodynamic,roovers1986linear,douglas1990characterization}) to $\Psi^{*}=20\sqrt{5}/(3^{5/2}\sqrt{\pi})\approx1.61859$ (hard-sphere value~\cite{douglas1990characterization,yamakawa1971modern}) for $f\to\infty$, where the polymer chains can be regarded as noninterpenetrating spheres.
Because $c^{*} \propto 1/(R_{g}^{3}\Psi^{*})$ is an increasing sequence of $f$, the decrease in $R_{g}^{3}$ is more dominant than the increase in $\Psi^{*}$ as $f$ increases.
The inset in Fig.~\ref{fig:previous}(b) demonstrates that $\Psi^*$ is a universal parameter depending only on $f$; $\Psi^{*}$ is independent of $M$, $T$, and the kind of polymer and solvent, at each $f$, by using four polymer-solvent systems [poly(styrene)/benzene, poly(styrene)/toluene, poly(butadiene)/cyclohexane, and poly(isoprene)/cyclohexane] at various $M$ and $T$~\cite{sato1987second,nakamura1991third,miyaki1978excluded,okumoto1998excluded,okumoto2000excluded,yamamoto1971more,fukuda1974solution,roovers1980hydrodynamic,roovers1986linear,douglas1990characterization,roovers1974preparation,bauer1989chain}.
(Detailed references are provided in SM, Sec.~S5.)

Regarding (ii), we investigate $M$ and $f$ dependencies of $\Pi$ in the semidilute regime ($c \gg c^*$).
Equation~(\ref{eq:semidilute}) can be rewritten as
\begin{equation}
\Pi = \frac{KRT}{M{c^*}^{\frac{1}{3\nu -1}}}c^{\frac{3\nu}{3\nu -1}}.
\label{eq:Pi}
\end{equation}
Because $c^*\propto M^{1-3\nu}$ for any $f$ [Fig.~\ref{fig:previous}(a)], Eq.~(\ref{eq:Pi}) indicates that $\Pi$ is independent of $M$ not only for linear polymers but also for star polymers [$M$ dependence in Fig.~\ref{fig:previous}(c)].
This result elucidates the $M$-independent scaling law for star polymer solutions $\Pi \propto c^{3\nu/(3\nu-1)}$ in Fig.~\ref{fig:osmotic}(c).
Moreover, because $c^{*}$ is an increasing sequence of $f$ (at fixed $M$ and $T$), Eq.~(\ref{eq:Pi}) indicates that $\Pi$ is a decreasing sequence of $f$ at fixed $c$ [$f$ dependence in Fig.~\ref{fig:previous}(c)].
Therefore, because $\Pi$ of star polymer solutions follows the universal EOS~(\ref{eq:EOS}), the $f$ dependence of $\Pi$ is originated from the $f$ dependence of $c^*$.
Note that the $f$ dependence of $c^*$ remains independent of $M$ [Fig.~\ref{fig:previous}(a)].
Thus, even when $M$ is sufficiently large and the central core proportion (branching points) to the arms becomes infinitely small, the branching effect on osmotic pressure remains [red characters in Fig.~\ref{fig:previous}(c)].\\

\textit{Concluding remarks}.\,---
We experimentally measured $\Pi$ of star polymers with up to eight arms in good solvents near and above the $c^*$ regime.
We found that the universal EOS given in Eq.~(\ref{fig:EOS}) (Fig.~\ref{fig:EOS}) describes $\Pi$ of these systems, with decreasing $\Pi$ [Fig.~\ref{fig:osmotic}(b)] and increasing $c^{*}$ [Fig.~\ref{fig:osmotic}(d)] as the arm number $f$ increased.
We ensured the universality of Eq.~(\ref{eq:EOS}) (i.e., the function $f_\Pi$) for various polymers and solvents, 
by showing that the scaling relations $R_{g} \propto M^{\nu}$, $\Psi^{*} \propto M^{0}$, and $c^{*} \propto M^{1-3\nu}$ were the same as those for the linear polymer solutions [Figs.~\ref{fig:Rg}(b), \ref{fig:previous}(b), and \ref{fig:previous}(a)].
We also showed that $R_{g}$, $\Psi^{*}$, and $c^{*}$ are decreasing, increasing, and increasing sequences of $f$, respectively.
The scaling relation $c^{*} \propto M^{1-3\nu}$ demonstrates the universality of EOS in star polymer solutions to be consistent with increasing $M$ and $f$, indicating that the $f$ dependence of $\Pi$ remains even when $M \to \infty$ in the semidilute regime [Fig.~\ref{fig:previous}(c)].

Our findings are beneficial for controlling osmotic pressure in the biomedical and pharmaceutical applications of star polymers~\cite{wu2015star}, such as drug delivery, gene delivery, surface modifiers, tissue engineering, and medical devices.
The universality of EOS in star polymer solutions can further provide an accurate determination of the osmotic pressure of branched polymers~\cite{burchard1999solution}, including dendrimers~\cite{Abbasi2014dendrimers} and polymer gels~\cite{nakagawa2022star}.

\medskip

\begin{acknowledgments}
This work was supported by the Japan Society for the Promotion of Science (JSPS) through the Grants-in-Aid for JSPS Research Fellows Grant No.~202214177 to T.Y., Scientific Research (B) Grant No.~22H01187 to N.S., Early Career Scientists Grant No.~19K14672 to N.S., Scientific Research (A) Grant No.~21H04688 to T.S., Transformative Research Area Grant No.~20H05733 to T.S., and MEXT Program Grant No.~JPMXP1122714694 to T.S.
This work was also supported by JST through CREST Grant No.~JPMJCR1992 and Moon-shot R\&D Grant No.~1125941 to T.S.
\end{acknowledgments}

\bibliographystyle{apsrev}

\clearpage
\widetext

\begin{center}
\textbf{\Large Supplemental Material for:\\
``Universality of Osmotic Equation of State in Star Polymer Solutions''}
\end{center}
\quad\\

\twocolumngrid

\setcounter{equation}{0}
\setcounter{figure}{0}
\setcounter{table}{0}
\makeatletter
 \@addtoreset{equation}{section}

\def\theequation{S\arabic{equation}}
\def\thefigure{S\arabic{figure}}

\setcounter{figure}{0}
\renewcommand{\thefigure}{S\arabic{figure}}
\renewcommand{\thetable}{S\arabic{table}}

\section{EFFECT OF END-FUNCTIONAL GROUPS ON OSMOTIC PRESSURE}

\begin{figure}[t!]
\centering
\includegraphics[width=\linewidth]{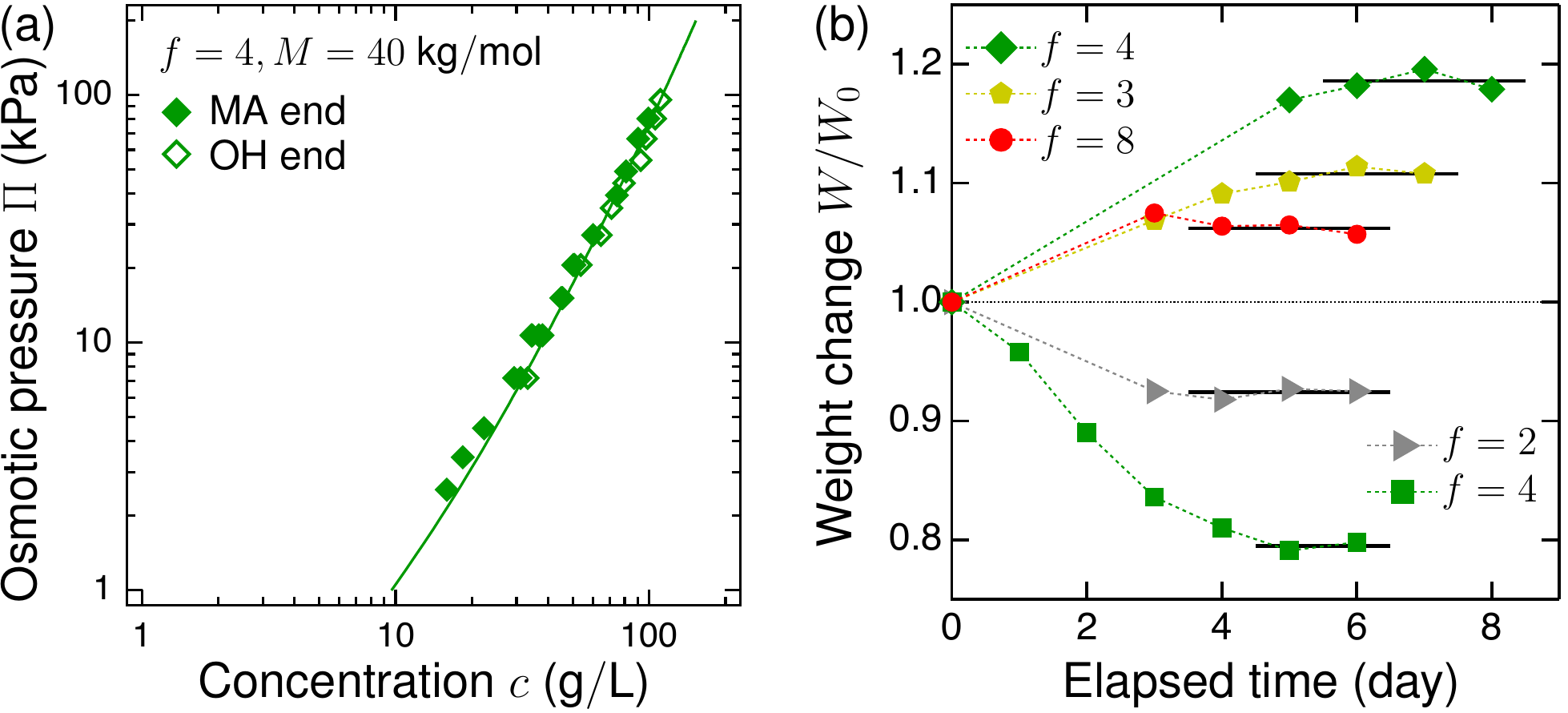}
\caption{
(a) Polymer mass concentration (c) dependence of the osmotic pressure ($\Pi$) in the four-arm star PEG solutions with different functional groups of MA (green-filled diamonds) and OH (green-open diamonds) with $M=40$ kg$/$mol.
The data of green-filled diamonds are the same as those in Figs.~1 and 2 in the main text.
(b) Time course of the weight-swelling ratio $W/W_{0}$ from the as-prepared state ($W_0$) to the equilibrium state ($W$) of PEG solutions in membrane osmometry.
For the linear PEG, we set $c_\mathrm{ext}=60$ g$/$L for $c_{0}=40$ g$/$L with $M=20$ kg$/$mol (gray triangles).
For the three-arm star PEG, we set $c_\mathrm{ext}=55$ g$/$L for $c_{0}=30$ g$/$L with $M=20$ kg$/$mol (yellow pentagons).
For the four-arm star PEG, we set $c_\mathrm{ext}=90$ g$/$L for $c_{0}=40$ g$/$L with $M=10$ kg$/$mol (green squares) and $c_\mathrm{ext}=80$ g$/$L for $c_{0}=60$ g$/$L with $M=40$ kg$/$mol (green diamonds).
For the eight-arm star PEG, we set $c_\mathrm{ext}=140$ g$/$L for $c_{0}=130$ g$/$L with $M=40$ kg$/$mol (red circles).
}
\label{fig:end}
\end{figure}

We demonstrate that the difference in the end-functional groups is negligible for the osmotic pressure $\Pi$.
Figure~\ref{fig:end}(a) shows $\Pi=\Pi(c)$ for four-arm star polymer (PEG) solutions with maleimide (MA, green-filled diamonds) and hydroxy (OH, green open diamonds) end-functional groups with $M=40$ kg$/$mol.
The variation in $\Pi$ is within experimental accuracy.
\\

\section{VERIFICATION OF REACHING EQUILIBRIUM}

To ensure that each solution sample reached equilibrium in membrane osmometry, we show the time course of the weight-swelling ratio $W/W_{0}$ for the linear, three-arm, four-arm, and eight-arm star polymer (PEG) solutions in Fig.~\ref{fig:end}(b).
Approximately one week was required to achieve an equilibrium state (horizontal black lines).
Thus, we determined the equilibrium state as the point at which $W/W_{0}$ remained constant for two to three days.
\\

\begin{table}[t!]
\label{table:0}
\caption{Data of the osmotic pressure ($\Pi$) for linear, three-arm, four-arm, and eight-arm star poly(ethylene glycol) in aqueous solutions at $298$~K measured in this study.}
\begin{ruledtabular}
\begin{tabular}{D{.}{.}{-1}D{.}{.}{3}D{.}{.}{-1}D{.}{.}{3}D{.}{.}{-1}D{.}{.}{3}}
\multicolumn{1}{c}{$c$} & 
\multicolumn{1}{c}{$\Pi$} & 
\multicolumn{1}{c}{$c$} & 
\multicolumn{1}{c}{$\Pi$} &
\multicolumn{1}{c}{$c$} & 
\multicolumn{1}{c}{$\Pi$}\\
\multicolumn{1}{c}{(kg$/$m$^{3}$)} & 
\multicolumn{1}{c}{(kPa)} & 
\multicolumn{1}{c}{(kg$/$m$^{3}$)} &
\multicolumn{1}{c}{(kPa)} &
\multicolumn{1}{c}{(kg$/$m$^{3}$)} &
\multicolumn{1}{c}{(kPa)}\\
\hline
\multicolumn{6}{c}{\textbf{Linear with $M=20$ kg$/$mol}}\\
16.4 & 3.45 & 45.3 & 27.18 & 73.3 & 66.70\\
18.4 & 4.51 & 54.0 & 34.98 & 82.8 & 80.38\\
20.0 & 5.76 & 55.5 & 34.98 & 83.3 & 54.64\\
22.6 & 7.20 & 57.8 & 44.10 & 83.6 & 80.38\\
30.7 & 10.71 & 58.2 & 44.10 & 95.9 & 95.74\\
37.9 & 15.15 & 69.7 & 66.70 & 97.5 & 112.99\\
39.2 & 15.15 & 70.3 & 54.64 & 99.9 & 95.74\\
43.6 & 20.60 & 72.1 & 54.64 & 109.2 & 112.99\\
\hline
\multicolumn{6}{c}{\textbf{Three-arm with $M=20$ kg$/$mol}}\\
19.7 & 5.76 & 45.3 & 20.60 & 76.7 & 54.64\\
24.3 & 7.20 & 45.8 & 20.60 & 86.2 & 66.70\\
27.0 & 8.85 & 56.2 & 27.18 & 90.1 & 80.38\\
33.4 & 12.81 & 69.2 & 34.98 & 96.5 & 95.77\\
40.2 & 15.15 & 73.7 & 44.10 &  & \\
\hline
\multicolumn{6}{c}{\textbf{Four-arm with $M=10$ kg$/$mol (Ref.~\cite{yasuda2020universal})}}\\
19.7 & 7.20 & 43.5 & 23.75 & 83.5 & 66.70\\
25.5 & 8.85 & 50.2 & 27.18 & 88.6 & 66.70\\
25.8 & 10.71 & 50.8 & 27.18 & 96.8 & 80.38\\
26.9 & 10.71 & 56.6 & 34.98 & 98.3 & 80.38\\
28.5 & 12.81 & 58.5 & 34.98 & 102.7 & 95.77\\
31.4 & 15.15 & 63.8 & 44.10 & 107.4 & 95.77\\
36.0 & 15.15 & 64.3 & 44.10 & 108.6 & 95.77\\
37.4 & 17.74 & 75.0 & 54.64 & 121.4 & 112.99\\
40.1 & 20.60 & 76.0 & 54.64 &  & \\
\hline
\multicolumn{6}{c}{\textbf{Four-arm with $M=40$ kg$/$mol (Ref.~\cite{yasuda2020universal})}}\\
15.9 & 2.55 & 34.6 & 10.71 & 60.1 & 27.18\\
18.4 & 3.45 & 37.9 & 10.71 & 74.5 & 39.37\\
22.3 & 4.51 & 45.0 & 15.15 & 81.0 & 49.18\\
29.3 & 7.20 & 45.5 & 15.15 & 90.4 & 60.70\\
31.1 & 7.20 & 50.1 & 20.60 & 99.1 & 80.38\\
34.3 & 10.71 & 50.7 & 20.60 &  & \\
\hline
\multicolumn{6}{c}{\textbf{Eight-arm with $M=40$ kg$/$mol}}\\
18.0 & 2.55 & 49.4 & 10.71 & 109.2 & 66.70\\
20.2 & 3.45 & 52.1 & 15.15 & 109.9 & 66.70\\
22.8 & 4.51 & 55.8 & 17.74 & 111.2 & 54.64\\
29.4 & 4.51 & 72.4 & 20.60 & 117.0 & 80.38\\
30.1 & 4.51 & 77.1 & 27.18 & 121.1 & 80.38\\
33.3 & 7.20 & 79.5 & 27.18 & 122.3 & 66.70\\
35.2 & 7.20 & 81.0 & 27.18 & 123.4 & 95.77\\
35.8 & 7.20 & 82.8 & 34.98 & 123.6 & 95.77\\
36.9 & 4.51 & 97.1 & 34.98 & 125.7 & 80.38\\
42.8 & 10.71 & 100.9 & 44.99 & 137.2 & 113.00\\
43.3 & 7.20 & 103.7 & 54.64 & 157.0 & 113.00\\
46.2 & 12.81 & 105.7 & 44.99 &  & \\
46.2 & 12.81 & 106.3 & 66.70 &  & \\
\end{tabular}
\end{ruledtabular}
\label{table:Pos}
\end{table}

\section{DATA OF OSMOTIC PRESSURE OF LINEAR AND STAR PEG IN AQUEOUS SOLUTIONS}

All experimental data of the osmotic pressure $\Pi$ at various polymer mass concentrations $c$ measured via membrane osmometry in this study are listed in Table.~\ref{table:Pos}.

\begin{figure}[t!]
\centering
\includegraphics[width=\linewidth]{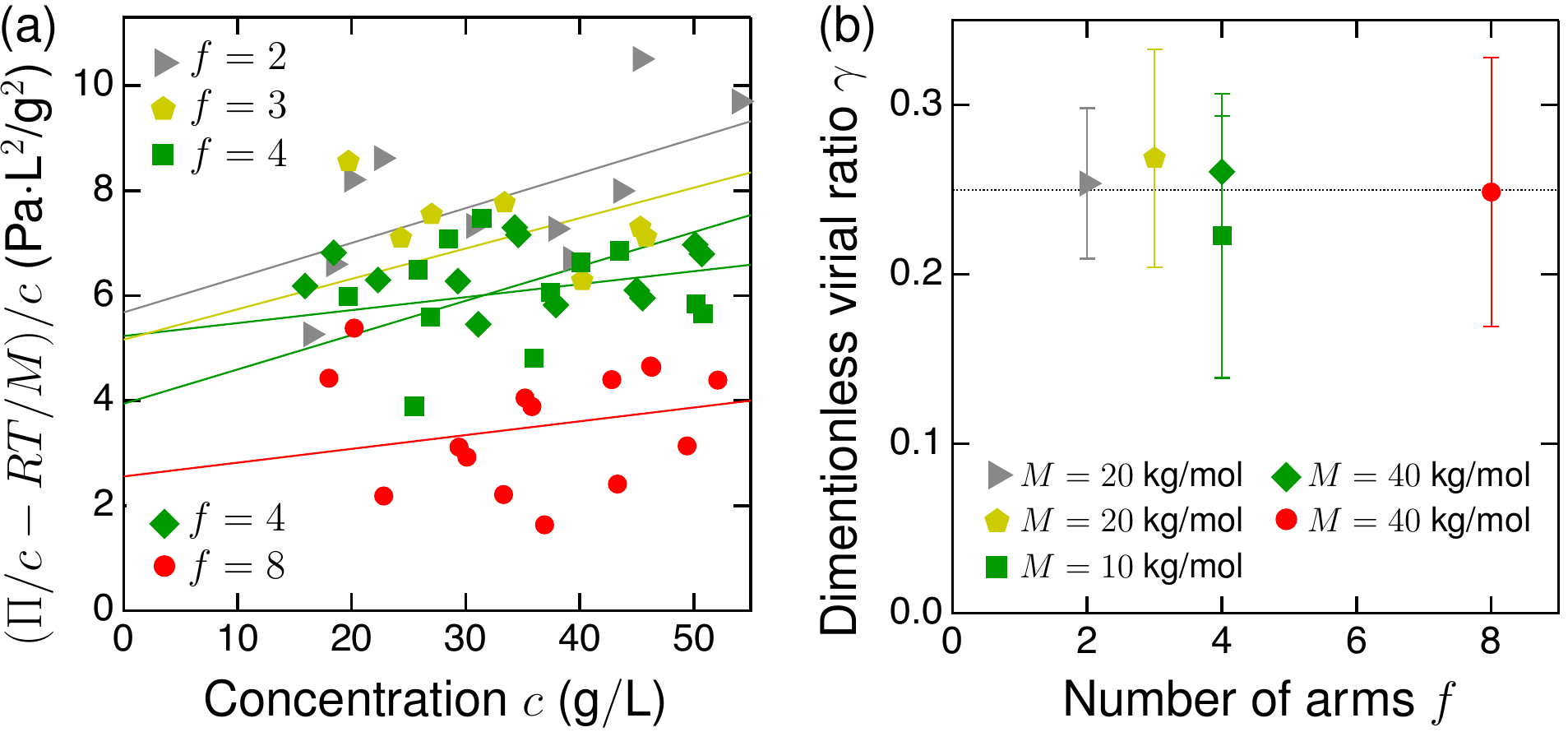}
\caption{(a) Evaluation of the dimensionless virial ratio $\gamma$.
Each solid line is obtained from the one-parameter least-square fit of $\gamma$ with Eq.~(\ref{eq:virial}) to the data for each sample, using $c^*$ estimated in Fig.~2(d) in the main text.
According to Eq.~(\ref{eq:virial}), the slope of each best fit line gives $\gamma$ of each arm number $f$.
(b) Dimension virial ratio $\gamma$ for various $f$.
The error bar indicates the standard error of the least-square fit.
The dashed line represents $\gamma\approx0.25$ for the linear polymer solutions~\cite{flory1953principles}.
}
\label{fig:third}
\end{figure}

\section{THIRD VIRIAL COEFFICIENT OF STAR POLYMER SOLUTIONS}

We evaluated the third virial coefficient in linear and star polymer (PEG) solutions based on the concentration ($c$) dependence of osmotic pressure $\Pi$.
In the universal EOS given by Eq.~(1) in the main text, the third virial coefficient $A_{3}$ corresponds to the dimensionless virial ratio $\gamma\equiv A_{3}/(A_{2}^{2}M)$ defined in Eq.~(2) in the main text.
From the virial expansion up to the third order [Eq.~(2) in the main text], we have 
\begin{equation}
\left( \frac{\Pi}{c} - \frac{RT}{M} \right) \frac{1}{c}= 
\frac{RT}{Mc^{*}}
\left(1 + \frac{c\gamma}{c^{*}}
\right).
\label{eq:virial}
\end{equation}
Using $M$ for each sample and $c^{*}$ evaluated in Fig.~2(d) in the main text at $298$~K, we show that the one-parameter least-squares fit to the data of $c$ dependence of $(\Pi/c - RT/M)/c$ in Fig.~\ref{fig:third}(a).
The slopes of the best-fit lines in Fig.~\ref{fig:third}(a) give $\gamma$ for each sample.
\\

For the linear, three-arm, four-arm, and eight-arm star polymer solutions, we have $\gamma=0.25(4)$, $\gamma=0.27(6)$, $\gamma=0.20(8)$ (with $M=10$ kg$/$mol) and $\gamma=0.26(3)$ (with $M=40$ kg$/$mol), and $\gamma=0.25(8)$, respectively.
Numbers in parentheses represent standard errors.
For the linear polymer solutions, the obtained $\gamma$ is consistent with $\gamma\approx0.25$~\cite{flory1953principles}.
Although the error bounds are not small for the three-, four-, and eight-arm star polymer solutions, the obtained $\gamma$ agrees well with $\gamma \approx 0.25$ for linear polymer solutions~\cite{flory1953principles}.
This result is consistent with the result that star polymer solutions with up to eight arms are described by the universal EOS~(Fig.~1 in the main text).
\\

\section{PREVIOUS RESEARCH FOR DILUTE LINEAR AND STAR POLYMERS IN GOOD SOLVENT}

In Figs.~3(b) and 4(a) and (b) in the main text, we exhibit the dilute solution properties of $R_{g}$, $A_{2}$, and $\Psi^*$ with various $M$, $T$, and types of polymer and solvent for linear and star polymers in good solvents.
All the data were measured through the light scattering reported in Refs.~\cite{sato1987second,nakamura1991third,miyaki1978excluded,okumoto1998excluded,okumoto2000excluded,yamamoto1971more,fukuda1974solution,roovers1980hydrodynamic,roovers1986linear,douglas1990characterization,roovers1974preparation,bauer1989chain,khasat1988dilute,huber1984dynamic,roovers1983analysis}.\\

For Figs.~2(b) and 3(a) in the main text, we used the 
linear (black upward triangles~\cite{sato1987second,nakamura1991third,miyaki1978excluded}), 
four-arm (green rightward triangles~\cite{okumoto1998excluded}), and 
six-arm (blue downward triangles~\cite{okumoto2000excluded}) 
poly(styrene) (PS) in benzene (Bz) at $298$~K.\\

For the main panel and the inset in Fig.~3(b) in the main text, we used data from 
linear PS in Bz at $298$~K (upward triangles~\cite{sato1987second,nakamura1991third,miyaki1978excluded}), 
linear PS in Bz at $303$~K (lower left triangles~\cite{fukuda1974solution} and upper left triangles~\cite{yamamoto1971more}), 
linear PS in toluene (TL) at $303$~K (black upper left triangles~\cite{yamamoto1971more}), 
linear PS in TL at $308$~K (squares~\cite{douglas1990characterization,roovers1980hydrodynamic}), 
linear poly(butadiene) (PB) in cyclohexane (CH) at $308$~K (squares~\cite{douglas1990characterization,roovers1986linear}), 
three-arm star PS in TL at $293$~K (lower-right triangles,~\cite{huber1984dynamic}), 
three-arm star PS in TL at $308$~K (upper-right triangles,~\cite{khasat1988dilute}), 
four-arm star PS in Bz at $298$~K (rightward triangles~\cite{okumoto1998excluded}), 
four-arm star PB in CH at $298$~K (squares~\cite{douglas1990characterization}), 
six-arm star PS in Bz at $298$ K (downward triangles~\cite{okumoto2000excluded}), 
six-arm star PB in CH at $298$~K (squares~\cite{douglas1990characterization,roovers1974preparation}),
eight-arm star poly(isoprene) (PI) in CH at $296$~K (diamonds~\cite{bauer1989chain}), 
ten-arm star PI in CH at $296$ K (diamonds~\cite{bauer1989chain}), 
twelve-arm star PI in CH at $296$ K (diamonds~\cite{bauer1989chain}), 
twelve-arm star PS in TL at $308$~K (leftward triangles~\cite{roovers1983analysis}), 
eighteen-arm star PB in CH at $298$~K (squares~\cite{douglas1990characterization}), 
eighteen-arm star PI in CH at $296$ K (diamonds~\cite{bauer1989chain}), 
eighteen-arm star PS in TL at $298$~K (leftward triangles~\cite{roovers1983analysis}), 
twenty-two arm star PI in CH at $296$ K (diamonds~\cite{bauer1989chain}), 
twenty-six arm star PI in CH at $296$~K (diamonds~\cite{bauer1989chain}), 
thirty-two arm star PI in CH at $296$~K (diamonds~\cite{bauer1989chain}), and 
fifty-six arm star PI in CH at $296$~K (diamonds~\cite{bauer1989chain}).

\end{document}